# "Open Innovation" and "Triple Helix" Models of Innovation:

# Can Synergy in Innovation Systems Be Measured?

Loet Leydesdorff [a] * & Inga Ivanova [b]

**Abstract**

The model of "Open Innovations" (OI) can be compared with the "Triple Helix of University-Industry-Government Relations" (TH) as attempts to find surplus value in bringing industrial innovation closer to public R&D. Whereas the firm is central in the model of OI, the TH adds multi-centeredness: in addition to firms, universities and (e.g., regional) governments can take leading roles in innovation eco-systems. In addition to the (transversal) technology transfer at each moment of time, one can focus on the dynamics in the feedback loops. Under specifiable conditions, feedback loops can be turned into feedforward ones that drive innovation eco-systems towards self-organization and the auto-catalytic generation of new options. The generation of options can be more important than historical realizations ("best practices") for the longer-term viability of knowledge-based innovation systems. A system without sufficient options, for example, is locked-in. The generation of redundancy—the Triple Helix indicator—can be used as a measure of unrealized but technologically feasible options given a historical configuration. Different coordination mechanisms (markets, policies, knowledge) provide different perspectives on the same information and thus generate redundancy. Increased redundancy not only stimulates innovation in an eco-system by reducing the prevailing uncertainty; it also enhances the synergy in and innovativeness of an innovation system.

[a] University of Amsterdam, Amsterdam School of Communication Research (ASCoR), PO Box 15793, 1001 NG Amsterdam, The Netherlands; email: loet@leydesdorff.net; * corresponding author
[b] Institute for Statistical Studies and Economics of Knowledge, National Research University Higher School of Economics (NRU HSE), 20 Myasnitskaya St., Moscow, 101000, Russia; and School of Economics and Management, Far Eastern Federal University, 8, Sukhanova St., Vladivostok 690990, Russia; inga.iva@mail.ru

**Introduction**

The model of "Open Innovations" (OI; Chesbrough, 2003) and the "Triple Helix of University-Industry-Government Relations" (TH; Etzkowitz & Leydesdorff, 1995 and 2000) seem at first sight to have much in common in terms of their stated objectives to bring industrial innovation closer to public R&D. On closer inspection, however, they differ in terms of their disciplinary backgrounds and policy objectives. As Chesbrough formulated (at p. xxiv), "Open innovation is a paradigm that assumes that firms can and should use external ideas as well as internal ideas, and internal and external paths to market, as the firms look to advance their technology." Firms are thus the principal agents. The TH focuses on the knowledge infrastructure of innovations provided by university-industry-government relations.

How can innovation eco-systems be improved? The transformation of the university toward an "entrepreneurial university" (Clark, 1998; Etzkowitz, 2002) and the role of innovation policies can be analyzed in terms of social coordination mechanisms that function differently from and beyond the market (Luhmann, 1995). From this perspective, the firm is one agent among others in networks of relations. In the OI model, however, the existing public/private divide in a political economy is more or less taken for granted, whereas the TH model calls attention to the newly emerging coordination mechanism of organized knowledge production in a knowledge-based economy (Leydesdorff, 2006; Whitley, 1999).

How is a knowledge-based economy different from a political economy? A political or industrial economy assumes markets and political institutions as the two most relevant coordination and selection mechanisms, while a knowledge-based economy is based on adding knowledge production as a third coordination mechanism to the mix. However, the generation of "wealth from knowledge" or "knowledge from wealth" requires knowledge-based mediation by management or intervention by government to change the institutional conditions (Freeman & Perez, 1988). The public/private divide is reconstructed in a knowledge-based economy, for example, in terms of intellectual property rights. A policy intervention such as the Bayh-Dole Act, for example, brings the industrial aspiration as a third mission into the core of the institutional arrangements between federal or national governments and national or state universities.

Arguing that applied research should be driven commercially and the university should not spend public money by investing in market development, the University of Amsterdam, for example, decided in the mid-90s to sell its science park to the Zernicke group (www.zernikegroup.com) which at that time was expanding internationally. In the meantime, however, the natural science faculty has been concentrated in a large park (of 70 ha), called the "Amsterdam Science Park." More than 120 (startup) companies have been incubated on this campus. However, in a recent study of the technology transfer performance of Dutch universities, Vinig & Lips (2015, at p. 1047) conclude that "(w)ith the exception of Dutch technical universities and academic medical centers, all Dutch research universities fail to translate their high research output into successful technology transfer and commercialization." While engaging in social relations, the traditional ("ivory-tower") university has become increasingly competent in shielding its intellectual research agendas against external interventions (van den Daele & Weingart, 1975). Incentives are considered as opportunities for funding.

**The neo-evolutionary turn of the TH**

How do these three selection environments (economic, scientific, and political) interact with one another? Selection environments operating upon one another reduce uncertainty (variation) potentially by orders of magnitude. Bruckner, Ebeling, Montaño, & Scharnhorst (1994) consider such markets as so "hyper-selective" that niches are needed to protect the incubation of innovation. Two layers of interactions are thus shaped: structural interactions among selection environments on top of ongoing interactions generating variation at the bottom level (e.g., new products and processes). The resulting system is vertically layered and horizontally differentiated. Because of this complex structure, the system can be considered nested (e.g., national, regional, etc.; Braczyk, Cooke, & Heidenreich, 1998) from one perspective, while clustered in overlapping sets from another (e.g., sectors; Breschi, Lissoni, & Malerba, 2003). Ivanova & Leydesdorff (2015) use the term "fractal manifold" to describe this fuzzy structure of innovation systems.

The two models—OI and TH—both depart from linear models such as "technology push" or "demand pull" in favor of a focus on interactions and further development. Relations are no longer fixed and given, as in a channel between a supply and a demand side (Kline & Rosenberg, 1986). The driving force in one phase may become a dependent variable in a next one; feedback and feedforward arrows co-determine longer-term development. A system with three interacting sources of variation cannot be expected to remain in equilibrium (Leydesdorff, 1994; Leydesdorff & Van den Besselaar, 1998). This "upsetting of the equilibria" (Nelson & Winter, 1982; Schumpeter [1939], 1964, p. 428) tends to become structural in a knowledge-based economy. Therefore, economic assumptions have to be reformulated in this neo-evolutionary framework (Andersen, 1994).

In the TH-model the relationships (among universities, industry, and government) were first conceptualized in terms of institutional relations, such as bilateral relations between universities and industries which may require administrative mediation or policy intervention. One can draw a triple helix as a triangle (Sábato, 1975) or as Venn diagrams (circles) that represent partly overlapping institutional spheres (Figure 1).

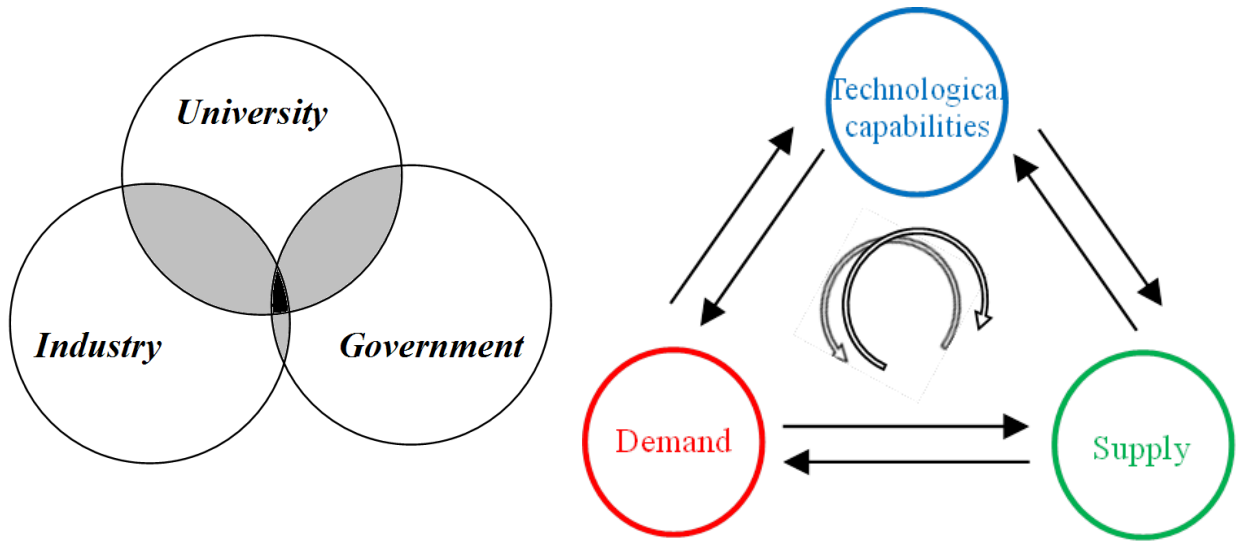

**Figure 1**: Representations of the Triple Helix in terms of a Venn diagram (Fig. 1a; source: Etzkowitz & Leydesdorff, 2000, p. 111) or as a triangle (Fig. 1b; source: Petersen *et al*., 2016, p. 667)

At the intersections in Figure 1a, one can pencil in relations such as funding of university research by government or industry, technology transfer, or priority programs formulated strategically, for example, at national levels. Institutional relations, however, tend to be sticky. For example, the Japanese government has promoted university-industry relations for decades, but individual researchers tend to prioritize the internationalization of their co-authorship relations more than their relevance at the national level (Leydesdorff & Sun, 2009). In other words, one can expect a tension between (*i*) integration and differentiation and (*ii*) the local and global dimensions. Korean firms, for example, are not always open to university initiatives because the knowledge base develops internationally. Knowledge-intensive firms (e.g., big pharma) tend to choose their university-partners globally, that is, in terms of the functionality of specific collaborations.

**Historical trajectories and evolutionary regimes**

In Figure 1b, agents and institutions are no longer in focus, but the TH model is generalized in terms of the three main functions in the innovation process: demand, supply, and technological capabilities. This model is no longer neo-institutional—that is, about networks of agents—but neo-evolutionary, since it is about the interactions among selection environments. How do markets interact with technologies in ways that are different from their interactions with institutions in a political economy? How are the institutional relations endogenously transformed by technological trajectories and emerging technological regimes—i.e., longitudinal selection mechanisms and their interactions?

Technological trajectories and regimes can be considered as a dually layered selection process (Dosi, 1982; Hayami & Ruttan, 1970; Nelson & Winter, 1982). A trajectory is historically observable along an "innovation avenue" (Sahal, 1985). The innovativeness of an eco-system, however, is also determined by absorptive capacities on the demand side and the skill structure of the labor force in the environment of the firm (Cooke & Leydesdorff, 2006). Technological innovation may make some natural resources (e.g., coal mines) obsolete and others (e.g., rare-earth metals) most valuable at the level of the global system.

Whereas each selection environment operates on specific variations, mutual selections may lead to co-evolution along trajectories in processes of "mutual shaping." In the case of three interacting selection environments, a technological regime can be expected additionally. This regime reorganizes the relevant variation beyond the control of the carrying agencies entertaining relations along historical trajectories. Each mutual relation (e.g., between "supply" and "demand") can be spuriously influenced by a third functionality pending as another possible selection context.

For example, highly industrialized countries and regions may be locked into dominant technologies or institutional arrangements and lose the flexibility to absorb new options. Are other arrangements possible? The neo-evolutionary version of the TH model considers the functions that are integrated and differentiated in the institutional relations. The institutional networks provide a knowledge-infrastructure, which from an evolutionary perspective can also be considered as a retention mechanism.

**Interactions among the helices**

The three main functionalities in the TH-triangle can be considered as (*i*) knowledge production (carried primarily by academia), (*ii*) wealth generation (industry), (*iii*) and normative control (governance). Since variations in these three dimensions can be considered as analytically independent, the three coordination mechanisms can be represented as orthogonal axes of a Cartesian coordinate system (Figure 2). Each agent or specific relation in an innovation system can be positioned in this space. From the perspective of the knowledge production system, for example, a university patent can be considered as relevant output; but patents are at the same time input to the economy. In the third dimension of governance and control, the patent may be filed under different national or international regimes such as USPTO, WIPO, EPO, etc.

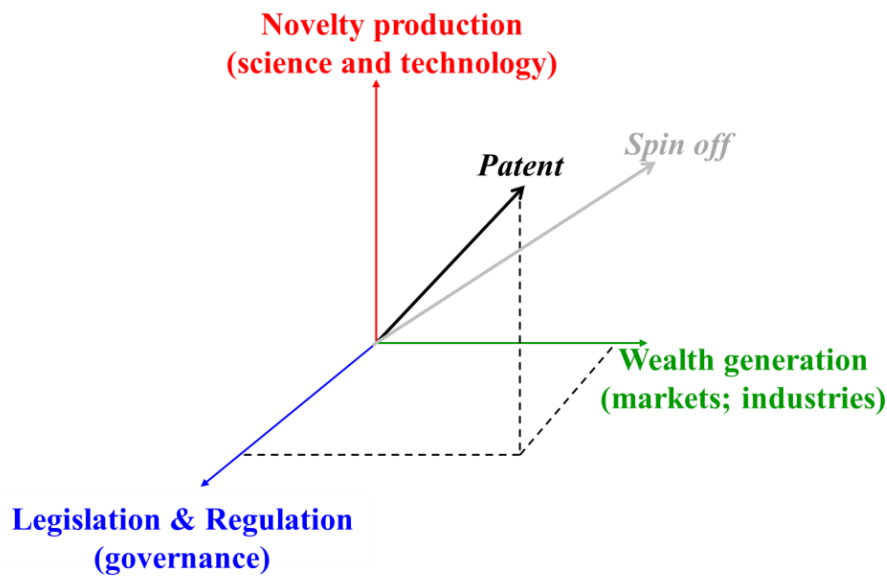

**Figure 2**: The Triple Helix as three functions in a system of Cartesian coordinates. Source: Ivanova & Leydesdorff, 2014a.

Unlike the institutions, the functions are not directly observable, but must be inferred as hypotheses by an analyst. The specification of these expectations shapes models that can be improved by systematic observation and thus exhibit a dynamics of knowledge-production that is different from market incentives or normative reasoning in political discourse (Hajer, Hoppe, Jennings, Fischer, & Forester, 1993). Because of the reflexivity involved, the knowledge dynamics can itself be expected to become part of the system that it models. The gradual transformation of a political economy into a knowledge-based economy can be expected to depend on the reflective capacities of this knowledge-based system: can the models be improved? Can other solutions be found? Has the loop added by scholarly reflections been self-reinforcing? Is a new specialty, for example, developing? (Rosenberg, 1982) The frame of reference is no longer the individual firms, but the knowledge-based reconstruction and transformation of the system of relations among innovation agents.

One can easily assume more than three dimensions to be relevant (Carayannis & Campbell, 2010; Leydesdorff, 2012). However, before we move to modeling an $n$-dimensional system, a three-dimensional one is worth further investigation: a TH can be expected to behave very differently from the sum of three double helices because the relations in a TH can loop forward or backward and thus generate fruition of or lock-in into an innovation eco-system (Ulanowicz, 2009). In other words, the third party may catalyze or inhibit the relations between the other two parties. Since each corner of the triangle can have this spurious function in relation to the other two, such a TH system can be considered as auto-catalytic or self-organizing.

The self-organization generates a next-order or global layer of communications on top of the local organization in terms of institutions and their relatively stable relations. This global layer (regime) can be considered as an order of expectations. The viability of options is first specified in terms of expectations. The specification of the expectations triggers and stimulates knowledge production processes to come to the fore when the discourse is further developed.

**Measurement**

The TH functionalities can be specified more abstractly (as latent dimensions) than the observable institutions and their relations. One moves from descriptive to inferential statistics when specifying expectations (before proceeding to the observations). In a recent paper, Petersen *et al.* (2016), for example, applied the TH methodology to the three main functions in the innovation process: (*i*) supply, (*ii*) demand, and (*iii*) technological capabilities (Figure 1b). In a series of studies of national systems of innovation, Leydesdorff and various co-authors modeled the TH as distributions of firms with (*i*) a geographical address (postal code), (*ii*) a technological knowledge base (using the NACE codes of the OECD as a proxy), and (*iii*) an economic weight (firm size). When observable distributions (variables) can be attributed to units of analysis—firms in this case—as independent (orthogonal) dimensions, the question can be raised of whether one would expect a synergy to emerge or would the system become locked into a vicious circle.[1]

---

[1] The model and the measurement enable us to specify this expectation for any three- or higher-dimensional system. One can find a routine for the computation of mutual redundancy in three or four dimensions at http://leydesdorff.net/software/th4/ .

Using firms as units of analysis in a series of studies, we decomposed a number of national systems of innovation: Germany (Leydesdorff & Fritsch, 2006), the Netherlands (Leydesdorff, Dolfsma, & van der Panne, 2006), Sweden (Leydesdorff & Strand, 2014), Norway (Strand & Leydesdorff, 2014), Italy (Cucco & Leydesdorff, manuscript), Hungary (Lengyel & Leydesdorff, 2011), the Russian Federation (Leydesdorff, Perevodchikov, & Uvarov, 2015), and China (Leydesdorff & Zhou, 2014). In the case of the Netherlands, Sweden, and China, the national level adds to the sum of the regions. In Sweden, the knowledge-based economy is heavily focused in three regions (Stockholm, Gothenburg, and Malmö/Lund); in China, four municipalities which are administered at the national level participate in the knowledge-based economy more than comparable regions.

In Norway, foreign-driven investment along the west coast seems to drive the transition from a political to a knowledge-based economy. Hungary's western part is transformed by the integration into the European Union, whereas the eastern part has remained a state-led innovation system. The capital Budapest occupies a separate position. In Germany, the generation of synergy is mainly at the level of the States (*Länder*) and not at the national level. In Italy, the main division is between the northern and southern parts of the country, and less so among regions as primarily administrative units. In the Russian Federation, the national level tends to disorganize synergy development at lower levels; knowledge-intensive services cannot circulate freely because of their integration in the Russian state apparatuses.

**Path-dependency, transition, regime change**

The transition from a dyad (supply/demand) to a triad (supply/demand/capability) is fundamental, as the sociologist Simmel already noted in 1902 (Simmel, 1950). A triad may be commutative or not: are the friends of my friends also my friends? The directionality of the arrows—the order of the communications—can generate asymmetries in triads: two loops in one direction and one in the other can be expected to lead to a path different from one loop in the first direction and two in the opposite. In other words, this system becomes path-dependent ("non-Abelian"): one cannot go back without friction to a previous state, as in an equilibrium system. In other words, a TH system is no longer in an equilibrium state, but necessarily in transition and developing (Etkzowitz & Leydesdorff, 1998).

Each point in the Cartesian space of Figure 2 can be considered as representing a three-dimensional vector in terms of its *x*, *y*, and *z*-coordinates. In the case of an event—and one expects events, since the system is developing—the corresponding vectors change. For example, University A may become more involved in industrial activities in the form of new startups. This can first be considered as a variant. If all or many universities move in this same direction, one would at a next moment have to rethink the choice of the axes in the vector space. For example, the axis of knowledge production could be rotated so that it points to the center of the cloud of points representing the universities.

In other words, a rotation of the structure is brought about by an aggregate of actions in a specific direction. This rotation can be clock-wise or counter-clockwise as illustrated in Figure 1b. Historical organization prevails in one direction and evolutionary self-organization in the other; but both organization and self-organization remain continuously relevant. In other words, historical organization and evolutionary self-organization are not an "either/or," but a question of extent or, in other words, a variable. This variable can be positive or negative (or zero); the question becomes one of measurement. How can one measure this variable?

Historically realized systems are measurable; but hypothesized systems are not yet necessarily realized. Information theory provides us with a language to express this: the realized options provide the observed uncertainty (in Shannon-type notation: $H_{obs}$), while our specifications of the system(s) provide us with an expectation of all possible states; that is, the maximum entropy $H_{max}$. The difference between the two ($H_{max}$ - $H_{obs}$) is non-information or redundancy $R$.[2] Redundancy is a measure of the options that could have been realized (given the definition of the system), but have not been realized hitherto. $R$ can be considered as the footprint of the next-order (possible) system in a historical configuration. For the viability of an innovation system, the availability of options other than the already realized ones may be more important than prior achievements. Redundancy is thus critical for innovation.

Because of Shannon's choice to couple the information measure $H$ to the entropy $S$,[3] the Second Law of thermodynamics is equally valid for $H$: entropy can only increase with time. In an

---

[2] Shannon (1948) defined the redundancy relative to the maximum information as follows: $R = (H_{max} - H_{obs})/H_{max}$.
[3] $H$ is a mathematical measure of uncertainty which Shannon (1948) coupled to the $H$ in Gibbs' formulation of the entropy: $S = k_B * H = k_B * - \sum_i p_i \log(p_i)$. In this formula, $k_B$ denotes the Boltzmann constant. When base 2 is used for the logarithm, $H$ is measured in bits, whereas $S$ (and $k_B$) are defined in Joule/Kelvin.

evolving system—such as a TH system—the $H_{max}$ can also be expected to increase with time. Brooks & Wiley (1986, at p. 43) have visualized this as follows (Figure 3a):

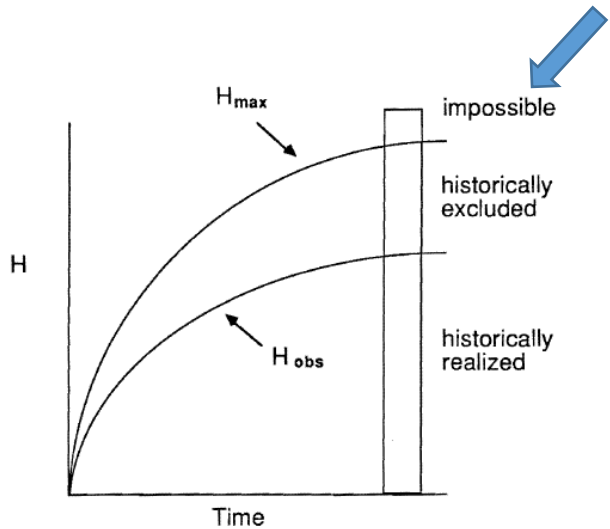 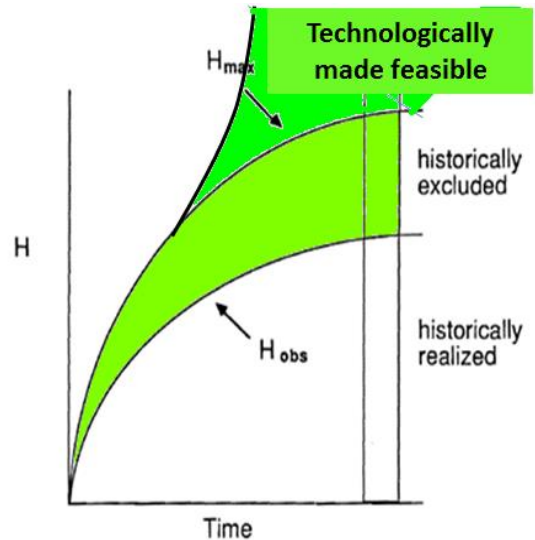

**Figure 3a**: The development of entropy ($H_{obs}$), maximum entropy ($H_{max}$), and redundancy ($H_{max} - H_{obs}$). Source: Brooks & Wiley (1986, at p. 43).

**Figure 3b:** Hitherto impossible options are made possible because of cultural and technological evolution. Source: Leydesdorff *et al.* (in press).

In other words, the generation of new options—that is, increase of redundancy—is at first a natural process. However, technological evolution adds to the redundancy by making the historically "impossible"—as indicated in the top-right corner of Figure 3a—feasible, and thus one adds non-natural (that is, humanly constructed) options to the system. We have added this domain in Figure 3b and colored the redundancy green. Note that more redundancy reduces uncertainty because the relative information ($H_{obs}/H_{max}$) is reduced. Reduction of uncertainty, for example, may shape niches in the complex system that are favorable to innovation more than when hyper-complexity and hyper-selectivity prevail.

**The generation of redundancy in TH systems**

How does one add redundancy (new options) to a system by producing knowledge? In addition to the functional dimensions represented as a vector space in Figure 2 above, one can consider the axes also as different perspectives on similar events: the academic perspective, the industrial one, and the political one. The overlaps in the Venn diagrams of Figure 1 in that case no longer indicate mutual information, but redundancy. One reads the same information, but from a different perspective.

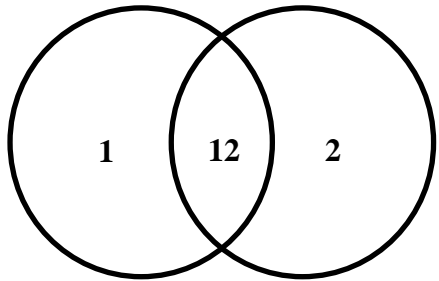

**Figure 4**: Overlapping uncertainties in two variables $x_1$ and $x_2$.

In Figure 4, the overlap between two variables $x_1$ and $x_2$ is depicted as two circles representing sets of values of each variable. The mutual information or transmission ($T_{12}$) is then defined—in accordance with the rules of set theory—as follows:

$$T_{12} = H_1 + H_2 - H_{12} \tag{1}$$

One corrects for the overlap by subtracting $H_{12}$. Alternatively, one can consider the overlap as redundancy: the same information is appreciated twice. In addition to $H_1$ and $H_2$, the overlap contains a surplus of information since both sides appreciate the overlap. This leads to an additional information as follows:

$$Y_{12} = H_1 + H_2 + T_{12} = H_{12} + 2T_{12} \tag{2}$$

The mutual redundancy $R_{12}$ at the interface between the two sets can now be found by using $Y_{12}$ instead of $H_{12}$ in Eq. 1, as follows:

$$\begin{aligned} R_{12} &= H_1 + H_2 - Y_{12} \\ &= H_1 + H_2 - (H_{12} + 2T_{12}) \\ &= H_1 + H_2 - ([H_1 + H_2 - T_{12}] + 2T_{12}) \\ &= -T_{12} \end{aligned} \tag{3}$$

Since $T_{12}$ is necessarily positive (Theil, 1972, pp. 59 ff.), it follows from Eq. 3 that $R_{12}$ is negative and *therefore* cannot be anything other than the consequence of an increased redundancy. Consequently, $R_{12}$ can be expressed in terms of negative amounts (e.g., bits) of information, that is, as reduction of uncertainty.

Leydesdorff, Petersen, and Ivanova (in press) derive in the case of more than two dimensions, *n* > 2:

$$R_n = (-1)^{1+n} T_{1234\ldots n} = [H(x_1, \ldots, x_n) - \sum_1^n H(x_i)]$$

$$+[\sum_{ij}^{\binom{n}{2}} T_{ij} - \sum_{ijk}^{\binom{n}{3}} T_{ijk} + \sum_{ijkl}^{\binom{n}{4}} T_{ijkl} - \cdots + (-1)^{1+n} \sum_{ijkl\ldots(n-1)}^{\binom{n}{n-1}} T_{ijkl\ldots(n-1)}] \quad (4)$$

The left-bracketed term of Eq. 4 is necessarily negative entropy (because of the subadditivity of the entropy), while the configuration of the remaining mutual information relations contribute a second term on the right which is positive.[4] In other words, we model here the generation of redundancy on the one side versus the historical process of uncertainty generation on the other, as an empirical balance in a system that operates with more than two codes (e.g., alphabets; Abramson, 1963, pp. 127 ff.). When the resulting $R$ is negative, self-organization prevails over organization in the configuration under study, whereas a positive $R$ indicates conversely a predominance of organization over self-organization as two different subdynamics.

**The multiplication of options in social systems**

Using biological or engineering metaphors, one often assumes that systems are "naturally" given and therefore have a maximum capacity. In other words, there are "limits to growth" (Club of Rome; Meadows, Meadows, & Randers, 1972) when the carrying capacity of a system is assumed to be "given" instead of specified in a scholarly discourse. For example, the capacity of transport across the Alps could be considered as constrained by the capacity of roads and railways such as at the Brenner Pass. As soon as one invents new channels, however, other options become available such as, for example, air transport across the Alps which is not constrained by the geological or weather conditions on the ground.

---

[4] The alternating sign in the right-hand term of Eq. 4 corrects for the alternating sign following from the Shannon formulas (Krippendorff, 2009, at p. 670).

The new options are not just added, but the number of options is multiplied with each new channel. Let us clarify this implication by using the following formalization: a network can be represented as a matrix ($n$ x $m$). Thus, its capacity is determined both by the number of units ($n$) and by the number of communications among these nodes ($m$). As long as $m$ is low, the aggregate number of units ($n$) can be used as an indicator of the capacity of the system. But with each increase of $m$, the influence of the number of units ($n$) on the product ($n$ x $m$) decreases.

For example, compare New York City with Calcutta in terms of "sustainability": living conditions in Calcutta are largely determined by the number of inhabitants because of the poor infrastructure. In New York, whether the city has ten or twelve million inhabitants does not particularly matter. The structure of this city is determined by its communication infrastructure (including such things as sewage systems, telephone lines, subways, etc.). A system which has changed the basis of its carrying capacity from actors to their interactive communications can grow exponentially. Each new column—representing another dimension of communication—*multiplies* the system's carrying capacity. However, the various columns represent also a *differentiation* of the communication into channels of communication. Thus, the number of options for innovation can rapidly increase if communication among the different codes of communication is appreciated in the model.

**Summary and conclusions**

Beyond opening the innovation process to third parties, the Triple Helix provides a model of innovation in which these third parties are specified in terms of selection environments and the interaction processes among them. Unlike the carrying layer in which innovation is understood to have developed historically, these different contexts provide meaning to innovation from specific perspectives. At the supra-individual level, these perspectives can be considered as differently coded communication systems. The codes are not present otherwise than as a structuration of the expectations (Giddens, 1984); yet the way and the extent to which these constructs interact matters for the innovation climate in terms of the numbers of options available for innovation and hence the reduction of uncertainty. These seemingly elusive processes can be modeled, measured, and simulated using the formulas that we submit (see also: Ivanova & Leydesdorff, 2014b).

Redundancy generation operates "against the arrow of time" or, in other words, in the direction opposite to the generation of Shannon-type information (that is, probabilistic entropy). In biological systems, pockets of "neg-entropy" (Brillouin, 1953; Schrödinger, 1944) can be expected as temporary niches in the entropy flux (Von Foerster, 1960). Differentiation among the codes of communication can be functional to the emergence of a cultural-technological evolution, which can tilt the balance between historical organization and evolutionary self-organization. Evolutionarily, the possible is no longer constrained by "adjacent others" in a historically defined universe (Kauffman, 2000), but by the quality of the models that one can entertain in the knowledge base (Luhmann, 1982). From a longer-term perspective, models of anticipatory systems (Dubois, 1998; Rosen, 1985) in which the arrow of time is also reversed, provide options for the simulation (Leydesdorff, 2010; Leydesdorff, Johnson, & Ivanova, 2014), while Triple-Helix models enable us to measure the efficiency of the anticipatory mechanisms that can be exploited for technological developments and innovations.

Both OI and TH models invite us programmatically to examine processes of exchange of information and knowledge among partners with different perspectives and other institutional roles. We have argued that exchanges at the level of the codes of communication can lead to redundancies that enrich the innovation process by making more and new options available. The richer context of a knowledge-based regime (at the supra-individual level) forces us in a feedback loop to specificity in the selections and to the knowledgeable legitimation of decisions. The realization of this additional loop stimulates the transition from a framework of politics (power) and economics (money) to one that contains organized knowledge production and innovation as a third mechanism of social coordination.